\documentclass[prd,aps,twocolumn,preprintnumbers,showpacs, nofootinbib,superscriptaddress,notitlepage]{revtex4-1}

\usepackage{slashed}
\usepackage{graphicx,color}
\usepackage{epsfig}
\usepackage{subfigure}
\usepackage{epsfig}
\usepackage{multirow}
\usepackage{epstopdf}
\usepackage{dcolumn}
\usepackage{bm}
\usepackage{color}
\usepackage{ulem}
\usepackage{enumitem}
\usepackage[colorlinks,
linkcolor=black,
anchorcolor=black,
citecolor=black
]{hyperref}

\begin{document}

	\baselineskip=15pt
	
	\preprint{CTPU-PTC-23-45}

\title{ Dark photon effects with the kinetic and mass mixing in Z boson decay processes}

\author{Jin Sun}
\email{sunjin0810@ibs.re.kr}
\affiliation{Particle Theory and Cosmology Group, Center for Theoretical Physics of the Universe, Institute for Basic Science (IBS), Daejeon 34126, Korea}
\author{Zhi-Peng Xing}
\email{zpxing@nnu.edu.cn(cooresponding author)}
\affiliation{Department of Physics and Institute of Theoretical Physics, Nanjing Normal University, Nanjing, Jiangsu 210023, China}

\begin{abstract}
Motivated by the most recent measurement of tau polarization in $Z\to \tau^+\tau^-$ by CMS, we have introduced a new $U(1)_X$ gauge boson field X, which can have renormalizable kinetic mixing with the standard model $U(1)_Y$ gauge boson field Y.
In addition to the kinetic mixing of the dark photon, denoted as $\sigma$, there may also be mass mixing introduced by the additional Higgs doublet with a vacuum expectation value (vev) participating in $U(1)_X$ and electroweak symmetry breaking simultaneously.
The interaction of the Z boson with the SM leptons is modified by the introduction of the mixing ratio parameter $\epsilon$, which quantifies the magnitude of both the mass and kinetic mixing of the dark photon.
Initially, we use the tau lepton as an example to explore the Z boson phenomenology of the dark photon model with both kinetic and mass mixing. The goal is to determine the allowed parameter regions by taking into account constraints from the vector and axial-vector couplings  $g_{V,A}^\tau$, the  decay branching ratio $Br(Z\to \tau^- \tau^+)$ and tau lepton polarization in $Z\to \tau^-\tau^+$. We found that   the mixing ratio plays important role in the Z boson features by choosing different  $\epsilon$ values.   
Furthermore, we aim to generalize our analysis from the tau-lepton case to include all fermions by conducting global fits. This allows us to identify viable regions by incorporating relevant fermion constraints and the W/Z mass ratio.
Correspondingly, we obtain the fit results with the kinetic mixing parameter $\sigma=0.074\pm0.021$, mixing ratio $\epsilon=-1.37\pm0.46$, and dark photon mass $m_X=275\pm39$ GeV. Our global analysis indicates a preference for a dark photon mass larger than $m_Z$.

\end{abstract}

\maketitle

\section{Introduction}

The study of beyond the standard model (BSM) has been the primary research direction in particle physics for the past few decades.
It has garnered significant attention, especially as an increasing number of new physics (NP) phenomena have been observed recently, such as the muon g-2~\cite{Muong-2:2023cdq} and the W mass deviation~\cite{Hays:2022qlw}, among others.
The explanation for these NP phenomena has sparked widespread interest in building BSM models.
Among the BSM models, one interesting example is the dark photon model. Recently, a systematic excess of single-electron events observed by the PHoton ELectron EXperiment (PHELEX)~\cite{dp} has further reinforced the credibility of the dark photon model.

In our work, we introduce a dark photon $X_\mu$ from the $U(1)_X$ gauge group, primarily coupled to a dark sector.
The dark photon can have kinetic mixing with the SM gauge bosons from $SU(3)_C \times SU(2)_L \times U(1)_Y$.
The kinetic mixing includes not only the abelian term $X^{\mu\nu} Y_{\mu\nu}$ with the kinetic mixing strength $\sigma$~\cite{Okun:1982xi,Galison:1983pa,Holdom:1985ag,Holdom:1990xp,Foot:1991kb}, 
 but also the non-abelian terms   $W^a_{\mu\nu} X^{\mu\nu}$ under the help of hypercharge-0 triplet scalar $\Sigma$~\cite{Fuyuto:2019vfe,Cheng:2021qbl}. 
 In addition to the simplest dark photon kinetic mixing, we further achieve new mass mixing by introducing additional Higgs doublets and singlets with dark charges. After the scalars obtain nonzero vacuum expectation values (vevs), they are responsible for breaking both the $U(1)_X$ and electroweak symmetries.
 The mass mixing pattern represents a novel dark photon mixing effect, which is clearly distinct from kinetic mixing, sometimes referred to as the dark Z model~\cite{Davoudiasl:2012ag}. The relevant phenomenology has been studied in Ref.~\cite{Davoudiasl:2012qa,Davoudiasl:2013aya, Davoudiasl:2014kua,San:2022uud,Goyal:2022vkg,Hosseini:2022urq,Biswas:2023uph}. 
 The two distinct mixing effects can be analyzed by defining the mixing ratio parameter $\epsilon$, which quantifies the magnitude of mass mixing and kinetic mixing. If the mixing ratio is zero ($\epsilon=0$), it corresponds to the typical kinetic mixing of dark photon. The mixing ratio plays a crucial role in phenomenological studies, such as using a smaller kinetic mixing parameter $\sigma$ to explain the CDF W mass results, etc~\cite{Cheng:2022aau,Davoudiasl:2023cnc,Yan:2022npz}.

 Recently, the CMS collaboration measured the latest result for the polarization of $\tau$ leptons in $Z\to \tau^+ \tau^-$ events, finding $P_\tau(Z)=-0.144\pm 0.015$~\cite{CMS:2023mgq}.
The result could be explained by the dark photon model with both kinetic mixing and mass mixing, as the polarization depends on the ratio of vector to axial-vector couplings. 
In our model, the interaction of the Z boson with SM fermions ($g_{V,A}^l$) in the neutral current expression will be modified by the two mixing patterns mentioned above, leading to an impact on the decay branching ratio of $Z\to l^- l^+$~\cite{pdg}. 
Initially, we concentrate on the $\tau$ lepton case to analyze the detailed phenomenology of the Z boson. Furthermore, we generalize our study to include all fermions, investigating the physical effects of mixing through global electroweak fits to determine suitable parameter ranges.

In this manuscript, we analyze the effects of the dark photon with both kinetic and mass mixing, aiming to identify viable parameter spaces that satisfy these relevant constraints simultaneously.

\section{The dark photon model with the kinetic and mass mixing }

The  gauge group of dark photon model is $SU(3)_C\times SU(2)_L\times U(1)_Y\times U(1)_X$. The SM fermions do not interact with the dark photon X from $U(1)_X$ because they do not carry the $U(1)_X$ charge. Fortunately, the dark photon X can have a kinetic mixing with  the hypercharge field from $U(1)_Y$. 
The kinetic terms of the bare fields $\tilde X'$ and $\tilde Y'$ and their interactions with other particles can be written as 
\begin{eqnarray}
	\mathcal{L} &=& -{1\over 4} \tilde X'_{\mu\nu} \tilde X'^{\mu\nu} - {\sigma \over 2} \tilde X'_{\mu\nu} \tilde Y'^{\mu\nu} - {1\over 4} \tilde Y'_{\mu\nu} \tilde Y'^{\mu\nu} \nonumber\\	&&+ j^\mu_Y \tilde Y'_\mu + j^\mu_X \tilde X'_\mu\;.
\end{eqnarray}
Here $j^\mu_X$ and $j^\mu_Y$ denote interaction currents of gauge fields $\tilde X'$ and $\tilde Y'$, respectively. The parameter $\sigma$ indicates the strength of the kinetic mixing.

To write the above Lagrangian in the canonical form one needs to diagonalize the kinetic terms
of X and Y.  There are  two commonly used ways of removing the mixing, namely, a) ~\cite{Holdom:1985ag, Dobrescu:2004wz}
the mixing term is removed in such a way that dark photon $\tilde X$ in the canonical form does not
couple to hyper-charge current $j_Y^\mu$, which is widely used to analyze the massless dark photon.
 And b)~\cite{Foot:1991kb, Babu:1997st} the hyper-charge field in the canonical form $\tilde Y$
does not couple to dark current $j^\mu_X$ produced by some dark particles with $U(1)_X$ charges, which is widely used in the studies of a massive dark photon. The two ways have been analyzed in detail in Ref.~\cite{Pan:2018dmu}.  Here the case b) is chosen to  conduct the analysis because we focus on the massive dark photon. 

The transformation for the case b) is 
\begin{eqnarray}\label{caseb}
	&&\left (\begin{array}{c}
	\tilde Y'\\\\
	\tilde X'
\end{array}
\right )
=
\left (\begin{array}{cc}
	1 \;&\;\;\;-\frac{\sigma}{\sqrt{1-\sigma^2}}\\\\
	0 \;&\;\;\; \frac{1}{\sqrt{1-\sigma^2}}
\end{array}
\right )
\left (\begin{array}{c}
	\tilde Y\\\\
	\tilde X
\end{array}
\right )\;,
\end{eqnarray}
By adopting Eq.~(\ref{caseb}), we can obtain 
the Lagrangian in the canonical form as
\begin{eqnarray}
	\mathcal{L} &=& -{1\over 4} \tilde X_{\mu\nu} \tilde X^{\mu\nu}   - {1\over 4} \tilde Y_{\mu\nu} \tilde Y^{\mu\nu} + j^\mu_Y \left(\tilde Y_\mu - \frac{\sigma}{\sqrt{1-\sigma^2}}\tilde X_\mu\right)\nonumber\\
	&&+ j^\mu_X \frac{1}{\sqrt{1-\sigma^2}}\tilde X_\mu\;.
\end{eqnarray}

The above analysis is actually the dark photon kinetic mixing case. For our model,  the main difference is scalar fields, which further affects symmetry breaking patterns.  The scalar fields include two Higgs doublet $\phi_1:(1,2,1/2)(0),  \; \phi_2:(1,2,1/2)(1)$ with $v_i$ (i=1,2) and one singlet $\phi_d:(1,1,0)(1)$ with vev $v_d$. Here $\sqrt{v_1^2+v_2^2}=v=246\mbox{GeV}$. The numbers in the bracket correspond to the quantum numbers in gauge group $SU(3)_C\times SU(2)_L\times U(1)_Y\times U(1)_X$.   Due to the $U(1)_X$ charge in $\phi_2$ and $\phi_d$, they will both contribute to  $U(1)_X$ symmetry breaking while $\phi_1, \phi_2$ broken the electroweak symmetry. This kind of broken pattern is obviously different from the typical dark photon kinetic mixing model.

The mass terms of gauge bosons are from the  scalar kinetic term given by
\begin{eqnarray}
	&&\mathcal{L}_{scalar} = \Sigma_i |D_\mu \phi_i|^2, \;\;\;\mbox{with} \nonumber\\
	&&D_\mu \phi_i=(\partial_\mu+ig' Y \tilde Y' _\mu+i g T_i \tilde W_{i\mu} +ig_X Q_d \tilde X'_{\mu})\phi_i\;.
\end{eqnarray}
Similarly, we need to adopt the transformation in Eq.~(\ref{caseb}) to express the fields in the canonical form $\tilde Y$ and $\tilde X$.    The relevant Higgs physics are analyzed in Ref.~\cite{Davoudiasl:2012ag}. In the following we mainly focus on the dark photon effects.

After electroweak symmetry breaking, $\tilde Y$ and the neutral component of the $SU(2)_L$ gauge field $\tilde W^3$ can be written in the combinations of the ordinary SM photon field $\tilde A$ and the $Z$ boson field $\tilde Z$ as follows
\begin{eqnarray}\label{SM}
	\tilde Y_\mu = c_W \tilde A_\mu -s_W \tilde Z_\mu\;,\;\;  W^3_\mu =  s_W \tilde A_\mu +  c_W \tilde Z_\mu\;,
\end{eqnarray}
where $c_W \equiv \cos \theta_W$ and $s_W \equiv \sin  \theta_W$ with $\theta_W$ being the weak mixing angle. 

Meanwhile, the $\tilde Z$ and $\tilde X$ fields both receive the mass $ m_Z$ and $ m_X$ as 
\begin{eqnarray}
		\mathcal{L}_{mass} &=& \frac{1}{2}(\tilde Z^\mu,  \; \tilde X^\mu) M^2  \left (\begin{array}{c}
			\tilde Z_\mu\\\\
			\tilde X_\mu
		\end{array}
		\right ),
\end{eqnarray}
The corresponding matrix element of mass squared are
 \begin{eqnarray}\label{mass matrix}
	M^2&=&\left (\begin{array}{cc}
		M_{11}^2\;\;\; & \;\;\; M_{12}^2\\\\
		M_{12}^2 \;\;\;&\;\;\;  M_{22}^2
	\end{array}
	\right )\nonumber\\
	&=&
	\left (\begin{array}{cc}
		m^2_{\tilde Z} \;\;\;&\;\;\;m^2_{\tilde Z}\frac{\sigma  s_W(1-\epsilon)}{\sqrt{1-\sigma^2}}\\\\
		m^2_{\tilde Z}\frac{\sigma  s_W(1-\epsilon)}{\sqrt{1-\sigma^2}} \;\;\;&\;\;\; m^2_{\tilde X}+ m_{\tilde Z}^2 \sigma^2  s_W^2\frac{1-2\epsilon}{1-\sigma^2}
	\end{array}
	\right )\;.
\end{eqnarray}
Here $m_{\tilde Z}^2= g_Z^2 v^2/4$,  $m_{\tilde X}^2= g_X^2(v_2^2+v_d^2)/(1-\sigma^2)$ and $g_Z=\sqrt{g'^2+g^2}$.  Note that the difference with Ref.~\cite{Cheng:2022aau} is that there is one additional mixing term $\tilde Z^\mu \tilde X_\mu$. Therefore we can not write the $\tilde Z$ boson and $\tilde X$ boson mass terms directly. 

The way by introducing the additional Higgs doublet results in a mass mixing  parameterized by  $2g_X v_2^2/(g_Z v^2)$, which is  dependent of the production of vev $v^2_2$ and  coupling constant $g_X$. This new mixing effect is termed as the mass mixing, which is obviously different from the kinetic mixing $\sigma$. 

In order to characterize the contribution from the two mixing effects, we define the mixing ratio  $\epsilon$  as
\begin{eqnarray}
	\epsilon= \frac{1}{\sigma s_W}\frac{2g_X}{g_Z   }\frac{v_2^2}{v^2}=\frac{1}{\sigma }\frac{2g_X}{g' }\frac{v_2^2}{v^2}\;,
\end{eqnarray}
The  parameter depends on the coupling ratio $g_X/g'$ and vev ratio $v_2/v$. If assuming $g_X=0$ or $v_2=0$, it results in $\epsilon=0$ showing no mass mixing effects. The ratio parameter shows clearly the proportion occupied by these two different mixing effects, 
 which is very important for the following Z boson phenomenological analysis.

In order to obtain the mass matrix in the physical basis, we need further diagonalize the mass matrix in Eq.~(\ref{mass matrix}) by  introducing the mixing angle as
\begin{eqnarray}
	&&\left (\begin{array}{c}
		Z\\
		X
	\end{array}
	\right )
	=
	\left (\begin{array}{cc}
		c_\theta\;\;\;&\;\;\; s_\theta\\
		-s_\theta\;\;\;&\;\;\;c_\theta
	\end{array}
	\right )
	\left (\begin{array}{c}
		\tilde Z\\
		\tilde X
	\end{array}
	\right )\;,
\end{eqnarray}
with $c_\theta = \cos\theta$, $s_\theta = \sin\theta$, and 
\begin{eqnarray}\label{mixing}
	\tan(2\theta) = {2M^2_{12} \over M_{11}^2- M_{22}^2 } \approx \frac{2\sigma s_W (1-\epsilon)}{1-\tilde r^2}\;.
\end{eqnarray}
Here $\tilde r^2= m_{\tilde X}^2/ m_{\tilde Z}^2$.  We assume $\epsilon$ is small and use the perturbative expansions in the second equation.
The diagonal masses  $ m_Z^2 $ and $ m^2_X$ corresponding to $Z$ and $X$ are given, respectively,  by
\begin{eqnarray}
	m^2_Z &=&  m^2_{\tilde Z}  c^2_\theta 
	+ \left(m^2_{\tilde X}+ m_{\tilde Z}^2 \sigma^2  s_W^2\frac{1-2\epsilon}{1-\sigma^2}\right) s^2_\theta \nonumber\\
	&&+ 2 s_\theta c_\theta m^2_{\tilde Z}\frac{\sigma  s_W(1-\epsilon)}{\sqrt{1-\sigma^2}}\;,\nonumber\\
 m^2_X &=&  \tilde m^2_Z  s^2_\theta + 
	 \left(m^2_{\tilde X}+ m_{\tilde Z}^2 \sigma^2  s_W^2\frac{1-2\epsilon}{1-\sigma^2}\right) c^2_\theta \nonumber\\&&- 2 s_\theta c_\theta m^2_{\tilde Z}\frac{\sigma  s_W(1-\epsilon)}{\sqrt{1-\sigma^2}}\;.
\end{eqnarray}

Combining the all transformations in Eqs.~(\ref{caseb}, \ref{SM},\ref{mixing}), we obtain the resulting form as
\begin{eqnarray}
	&&\left (\begin{array}{l}
		\tilde A'\\\\
		\tilde Z'\\\\
		\tilde X'
	\end{array}
	\right )
	=\left ( \begin{array}{ccc}
		1\;&\; -s_\theta \frac{\sigma c_W}{\sqrt{1-\sigma^2}} \;&\;-c_\theta \frac{\sigma  c_W}{ \sqrt{1-\sigma^2  }}\\\\
		0\;&\; c_\theta+s_\theta \frac{\sigma s_W}{\sqrt{1-\sigma^2 }} \;&\; -s_\theta +c_\theta \frac{\sigma s_W}{\sqrt{1-\sigma^2}}\\\\
		0\;&\;s_\theta\frac{1}{\sqrt{1-\sigma^2}} \;&\;c_\theta{1\over \sqrt{1-\sigma^2 }}
	\end{array}
	\right )
	\left (\begin{array}{l}
		A\\\\
		 Z\\\\
		 X
	\end{array}
	\right )\;,
\end{eqnarray}

The general Lagrangian that describes the physical fields (A, Z, X)  kinetic energy, and their interactions with the electromagnetic  current $j^\mu_{em}$, neutral $Z$-boson current $j^\mu_Z$ and dark current $j^\mu_X$ is given by
\begin{eqnarray}\label{interaction}
	\mathcal{L} &= &-{1\over 4}  X_{\mu\nu}  X^{ \mu\nu} - {1\over 4}  A_{\mu\nu} A^{ \mu\nu} - {1\over 4}  Z_{\mu\nu}  Z^{ \mu\nu} 
	 \nonumber\\
	 && + j^\mu_{em} \left( A_\mu   - s_\theta  {\sigma  c_W \over \sqrt{1-\sigma^2} } Z_\mu -c_\theta{\sigma  c_W\over  \sqrt{1-\sigma^2 }}X_\mu \right)\nonumber\\
	&& + j^\mu_Z\left( \left( c_\theta+ s_\theta{\sigma s_W\over \sqrt{1-\sigma^2}}   \right )Z_\mu+ \left(-s_\theta+ c_\theta {\sigma s_W \over \sqrt{1-\sigma^2}}\right) X^\mu\right)\nonumber\\
	&&+j^\mu_X\left(  s_\theta{1\over \sqrt{1-\sigma^2}}Z_\mu  +c_\theta {1\over \sqrt{1-\sigma^2  }}  X_\mu \right).\label{darkzc}
\end{eqnarray}
 Here the currents for fermions with charge $Q_f$ and weak isospin $I^f_3$ in the SM are given by
\begin{eqnarray}
	&&j^\mu_{em} = - \sum_f  e Q_f \bar f\gamma^\mu f\;,\;\;j^\mu_Z = -{ e \over 2  s_W  c_W} \bar f\gamma^\mu ( \tilde g^f_V - \tilde g^f_A \gamma_5) f\;,\nonumber\\
	&& \tilde g^f_V = I^f_3 - 2 Q_f  s^2_W\;,\;\;\;\; \tilde g^f_A = I^f_3\;.
\end{eqnarray}
Note that the W boson field and its interactions are not affected directly.

Analyzing the SM interaction in Eq.~(\ref{interaction}), we found two interesting feutures. One is that  A is already the physical massless photon field $A$ without the need of further mass diagonalization. And the Z boson interaction will be affected by modifying the vector and axial-vector coupling constants as 
\begin{eqnarray}
	g^f_V &=& \tilde g_V^f \left(c_\theta+s_\theta { \sigma s_W \over \sqrt{1-\sigma^2}}\right)+2  s_\theta{\sigma  s_W  c_W^2 \over {\sqrt{1-\sigma^2}}}\nonumber\\
	&\approx&   \tilde g^f_V\left[ 1+ \sigma^2 \tilde s_W^2\left({1-\epsilon \over 1-\tilde r^2}  - {(1-\epsilon)^2 \over 2(1-\tilde r^2)^2}\right)\right]
	\nonumber\\&&+2\sigma^2  s_W^2 (1- s_W^2)\frac{1-\epsilon}{1-\tilde r^2}\;,\;\;\;\; \nonumber\\
	g^f_A &=&\tilde g^f_A \left(c_\theta+s_\theta { \sigma s_W \over \sqrt{1-\sigma^2}}\right)\nonumber\\
	&\approx&  \tilde g^f_A  \left[ 1+  \sigma^2 s_W^2 \left(-\frac{(1-\epsilon)^2}{2(1-\tilde r^2)^2}  +\frac{1-\epsilon}{1-\tilde r^2} \right) \right].
\end{eqnarray}

The above modification for  coupling  will affect the Z boson features with SM fermions. Analyzing the couplings,  we found that if $\epsilon=1$, the mass matrix in Eq.~(\ref{mass matrix}) has been a diagonal one without further diagonalization by introducing the mixing angle $\theta$. In this case,  the  kinetic  mixing effect for Z boson will be fully canceled out by  the  mass mixing so that dark photon has no contribution to SM fermions. And  the value of $\sigma$ will be influenced by the dark coupling $g_X$ and vev $v_2$. The limit $\epsilon=1$ corresponds with 
\begin{eqnarray}\label{ratio}
	\epsilon=1 \rightarrow \frac{2g_X}{g' }\frac{v_2^2}{v^2}=\sigma\;.
\end{eqnarray}
This means that if we chose $\sigma \sim 2\times 10^{-4} $,     $v_2/v \sim 0.01$ with $g_X \sim g'$ or  $g_X \sim 10^{-4}g'$ with $v_2/v \sim 1/\sqrt{2}$. 

Since the $\epsilon=1$ case is very special and trivial for analyzing this model, in the following we will focus on the dark photon mixing effects on Z boson coupling with SM fermions, which means $\epsilon \neq 1$. Based on Eq.~(\ref{ratio}), the mixing ratio  could be enlarged around many times by improving $g_X$ or $v_2$.  However, in order to satisfy the perturbative expansions for small mixing angle $\theta$  in Eq.~(\ref{mixing}), we keep $|1-\epsilon| \sim 10$ which corresponds to $\epsilon \in (-9,11)$.  This shows we could use the smaller kinetic mixing $\sigma$ to explain the CDF W mass results compared to $\sigma>0.15$ in Ref.~\cite{Cheng:2022aau}.  The detailed analysis can refer to Ref.~\cite{Davoudiasl:2023cnc}. 
Therefore, in the following   $Z$ boson phenomenological analysis,  we mainly focus on the range $\epsilon \in (-9,11)$.


\section{Tau-lepton phenomenology analysis}

In this part we will study the Z boson interaction with SM fermions, especially for tau lepton. For the tau lepton, the original vector and axial-vector couplings in SM are $\tilde g_V^\tau=-1/2+2s_W^2$ and $\tilde g_A^\tau=-1/2$.  Analyzing the constraints from the vector and axial-vector Z boson coupling with tau, we obtain the viable parameter space. Furthermore, the two couplings will influence the branching ratio and tau lepton polarization   in the Z boson decaying into tau lepton pairs $Z\to \tau^-\tau^+$.\\

\noindent
{\bf Vector and axial-vector couplings of Z boson with $\tau$ lepton }

For tau lepton,  the  vector coupling $ g_V^\tau$ and axial-vector couplings  $ g_A^\tau$  are given by 
\begin{eqnarray}
 g^f_V  &=&
	  \left(- \frac{1}{2}+2s_W^2\right)\left[ 1+ \sigma^2  s_W^2\left({1-\epsilon \over 1-\tilde r^2}  - {(1-\epsilon)^2 \over 2(1-\tilde r^2)^2}\right)\right]
	  \nonumber\\&&+2\sigma^2  s_W^2 (1- s_W^2)\frac{1-\epsilon}{1-\tilde r^2}\;,\;\;\;\; \nonumber\\
	  g^f_A &=&-\frac{1}{2}
	    \left[ 1+ \sigma^2  s_W^2\left(\frac{1-\epsilon}{  1-\tilde r^2 } -\frac {(1-\epsilon)^2}{ 2(1-\tilde r^2)^2}\right)\right]\;.
\end{eqnarray}
We found that the two couplings highly depends on the mixing ratio $\epsilon$.
 If  $\epsilon=1$, it returns to the SM case without dark photon mixing effects.
 If $\epsilon=0$, this corresponds with the dark photon model with only kinetic mixing.

 \begin{figure*}[!t]
 	\centering
 	\subfigure[\label{vector}]
 	{\includegraphics[width=.486\textwidth]{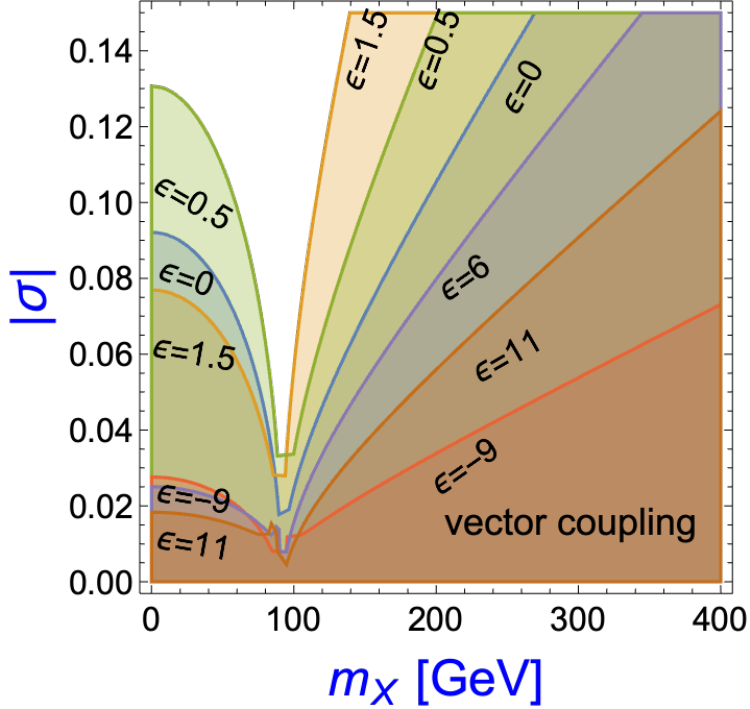}}
 	\subfigure[\label{axialvector}]
 	{\includegraphics[width=.486\textwidth]{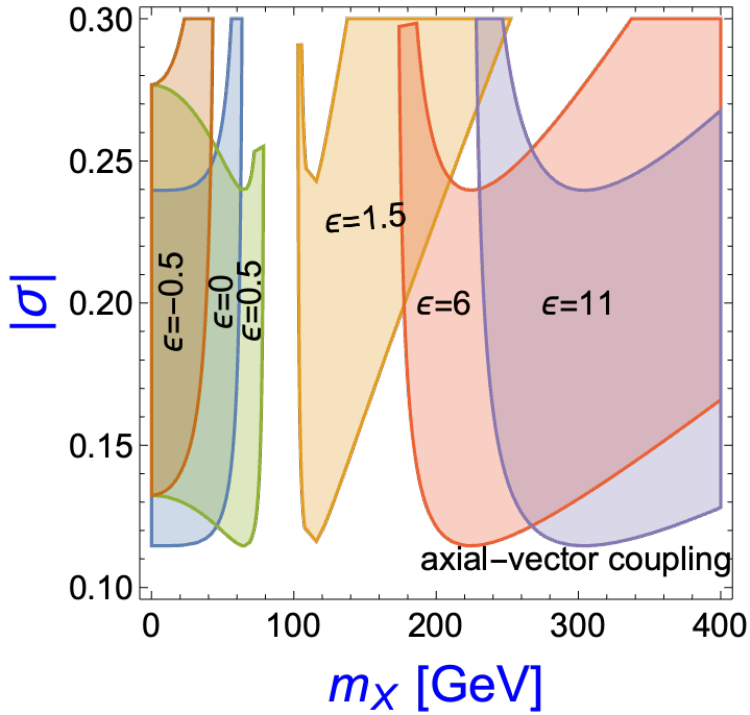}}
 	\caption{ The allowed parameter regions in $m_X-|\sigma|$ by vector and axial-vector couplings within $2\sigma$, respectively.  The different mixing ratio $\epsilon$ are shown in different color regions.
 	}
 	\label{coupling}
 \end{figure*}

Based on the current experimental values $g_V^\tau=-0.0366\pm0.0010$ and $	g_A^\tau=-0.50204\pm0.00064$~\cite{pdg}, we can obtain the allowed parameter ranges  within $2\sigma$ as shown in Fig.~\ref{coupling}.  The different mixing ratio corresponds with the different color region.  Here we choose the only kinetic mixing $\epsilon=0$,  reduced mixing $\epsilon=0.5$ and $\epsilon=1.5$,  enhanced mixing $\epsilon=6$,  the upper bound $\epsilon=11$ and lower bound $\epsilon=-9$.
The left panel shows the plane $m_X-|\sigma|$ with different mixing ratio for vector coupling. We found that  the allowed region will drop (rise) gradually with $\epsilon>1 \;(\epsilon<1)$ when increasing the ratio parameter $\epsilon$.  And the  regions cover the entire dark photon mass range, whatever  $m_X<m_Z$ or  $m_X>m_Z$.   The right panel corresponds with the case of  axial-vector coupling, which has fully different variation trend when changing the ratio parameter $\epsilon$. We find that when  increasing  $\epsilon$,  the allowed region will move right constantly. And  the region with $\epsilon>1\; (\epsilon<1)$ only exists in  $m_X>m_Z\; (m_X<m_Z)$. The  feature different  from the vector case shows the need of using two different mass range to conduct the analysis.  According to our further study, we also found  the axial-vector constraints have already excluded the case $\epsilon<-1$.
\\

\noindent
{\bf Z boson decaying into tau pairs $Z\to \tau^-\tau^+$}

The decay branching ratio  for $Z \to \tau^-\tau^+$ is given by
\begin{eqnarray}
	Br(Z\to \tau^+ \tau^-)=\frac{G_F m_Z^3}{6\sqrt{2}\pi \Gamma_Z}\sqrt{1-\frac{4m_\tau^2}{m_Z^2}}\left((g_V^{\tau })^2+(g_A^{\tau })^2\right).\nonumber\\
\end{eqnarray}
Here $G_F=1.1664\times 10^{-5} \mbox{GeV}^{-2}$ is the Fermi constant, $m_\tau$ is tau lepton mass and $\Gamma_Z=2.4952$ GeV~\cite{pdg}. 

\begin{figure*}[!t]
	\centering
	\subfigure[\label{branching}]
	{\includegraphics[width=.486\textwidth]{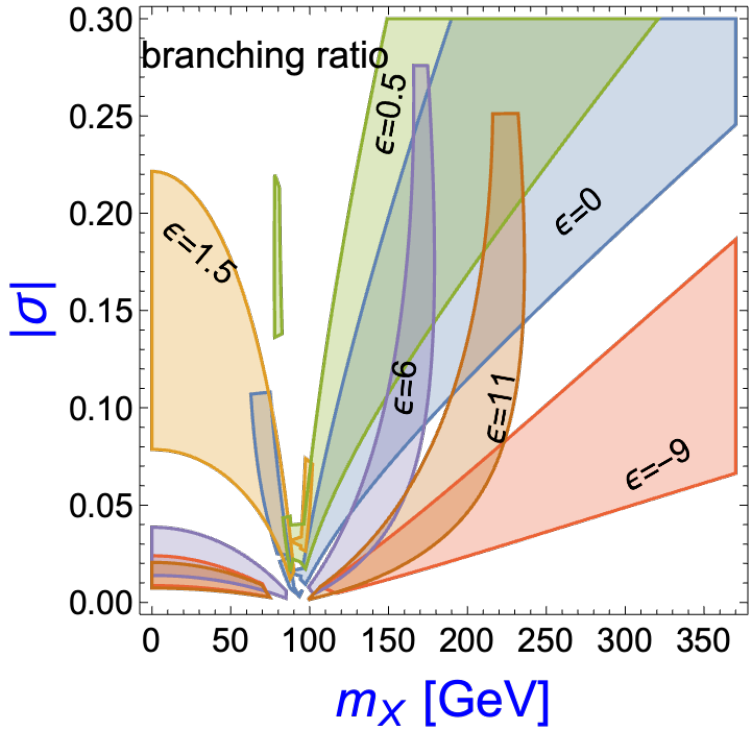}}
	\subfigure[\label{polarization}]
	{\includegraphics[width=.486\textwidth]{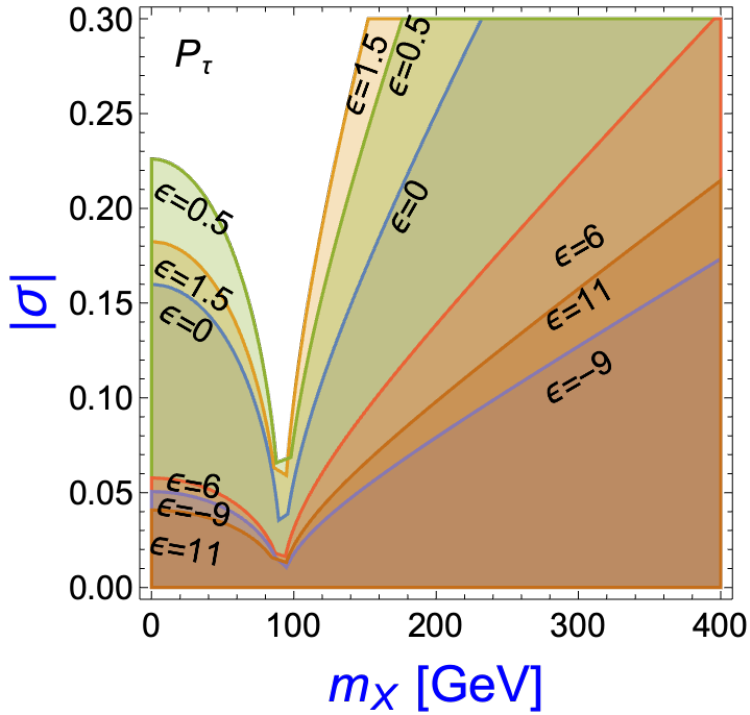}}
	\caption{ The allowed parameter regions in $m_X-|\sigma|$ by branching ratio and tau polarization for $Z\to \tau^-\tau^+$ within $2\sigma$, respectively.  The different mixing ratio $\epsilon$ are shown in different color regions.
	}
	\label{decay}
\end{figure*}

The branching ratio strongly depends on vector and axial-vector couplings.  Modifying the two couplings by the mixing effects will change the branching ratio correspondingly.  Using the current experimental values $Br(Z\to \tau^+ \tau^-)=(3.3696\pm 0.0083)\%$~\cite{pdg},    the allowed parameter regions within $2\sigma$ are shown in Fig.~\ref{branching}. Similarly, the different mixing ratio corresponds with the different color region.  We found that the allowed region in the $m_X-|\sigma|$ plane could both exist in $m_X<m_Z$ and  $m_X>m_Z$ region. For the $m_X<m_Z$ region,  increasing  $\epsilon$ will uplift (drop) the region with  $\epsilon<1\;(\epsilon>1)$.  For the  $m_X>m_Z$ region, increasing  $\epsilon$ will make the region move upper left ( right ) for $\epsilon<1 \;(\epsilon>1)$.

Besides the above branching ratio for $Z\to \tau^-\tau^+$,  another very important physical measurement quantity is the polarization of final tau lepton.  It is defined as 
\begin{eqnarray}
	P_\tau=\frac{\sigma_+(Z\to \tau^-\tau^+) - \sigma_- (Z\to \tau^-\tau^+)}{\sigma_+(Z\to \tau^-\tau^+) + \sigma_- (Z\to \tau^-\tau^+)} \;.
\end{eqnarray}
Here $\sigma_{\pm}$ shows the cross sections for the tau leptons with positive and negative  helicities, respectively.  The helicity of $\tau$ leptons from Z boson decays can be measured from the energy and angular distributions of the tau lepton decay products.   If the center of mass energy  is equal to the Z boson mass $m_Z$,  the polarization can be determined by the vector and axial-vector couplings $g_{V,A}^\tau$ for the tau lepton as
\begin{eqnarray}
	P_\tau&&=-2\frac{g_V^\tau  g_A^\tau}{\left((g_V^{\tau })^2+(g_A^{\tau })^2\right)}\nonumber\\
	&& \approx
	-2\left[1-4s_W^2- 4\sigma^2 s_W^2 c_W^2 \frac{1-\epsilon}{1-r^2}\right]\;.
\end{eqnarray}
Recently, the polarization is measured using leptonic and hadronic $\tau$ lepton decays in $Z\to \tau^-\tau^+$ events in proton-proton collisions at $\sqrt{s}=13$ TeV recorded by CMS with an integrated luminosity of $36.3 fb^{-1}$~\cite{CMS:2023mgq}. Using the experimental value $P_\tau(Z)=-0.144\pm0.015$, 
we can obtain the allowed parameter region within $2\sigma$ are shown in Fig.~\ref{polarization}.
We found that the region with different $\epsilon$ cover the entire dark photon mass range. The region of tau polarization has the same shape with vector coupling case, and the only difference is the magnitude of the kinetic mixing $\sigma$.  This is due to $P_\tau \approx -2g_V^\tau/g_A^\tau$ in the limit of $g_V^\tau<<g_A^\tau$. And the allowed region will drop (rise) gradually with $\epsilon>1 \;(\epsilon<1)$ when increasing the ratio parameter $\epsilon$.  \\

\noindent
{\bf Combining the above constraints  in $Z\to \tau^-\tau^+$}

\begin{figure*}[!t]
	\centering
	\subfigure[\label{mx50}]
	{\includegraphics[width=.486\textwidth]{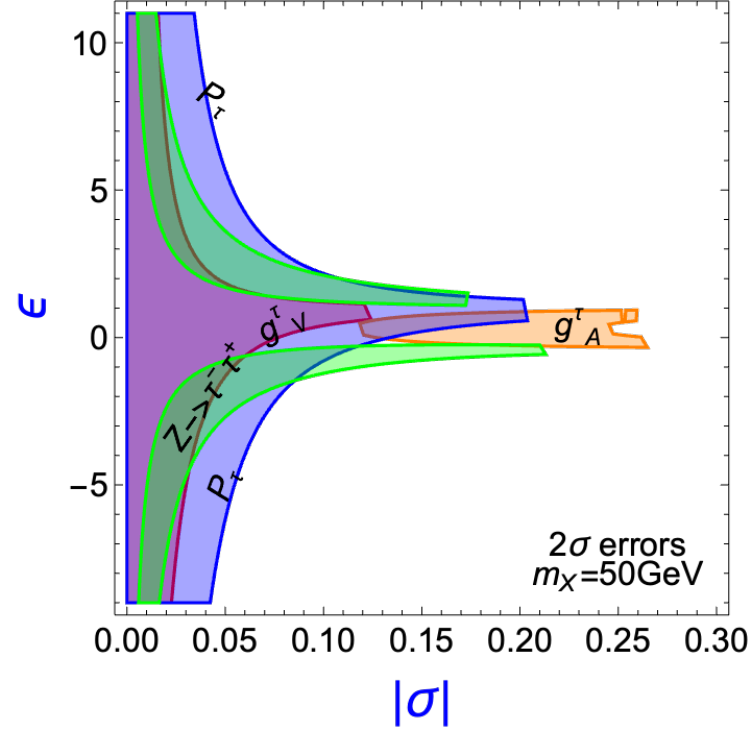}}
	\subfigure[\label{mx250}]
	{\includegraphics[width=.486\textwidth]{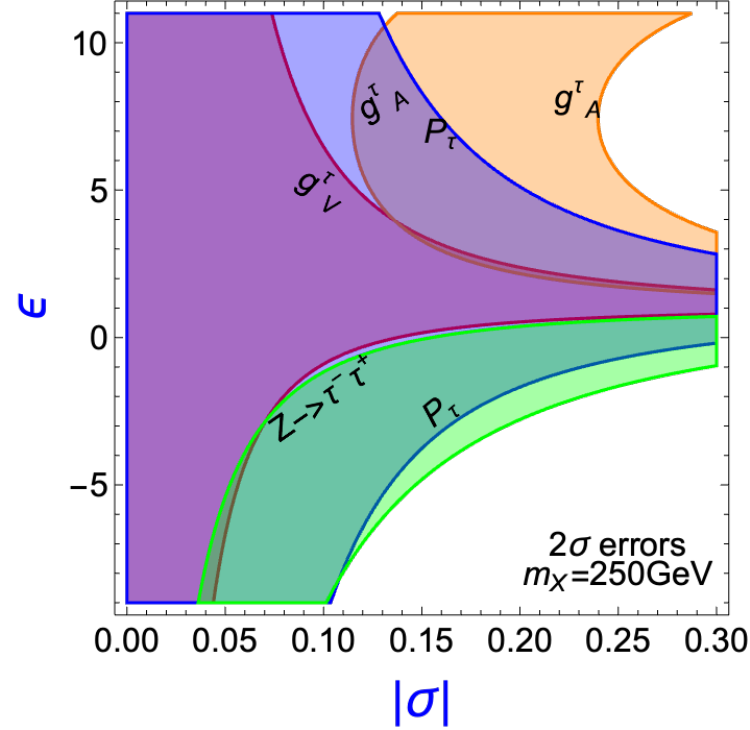}}
	\subfigure[\label{3mx50}]
	{\includegraphics[width=.486\textwidth]{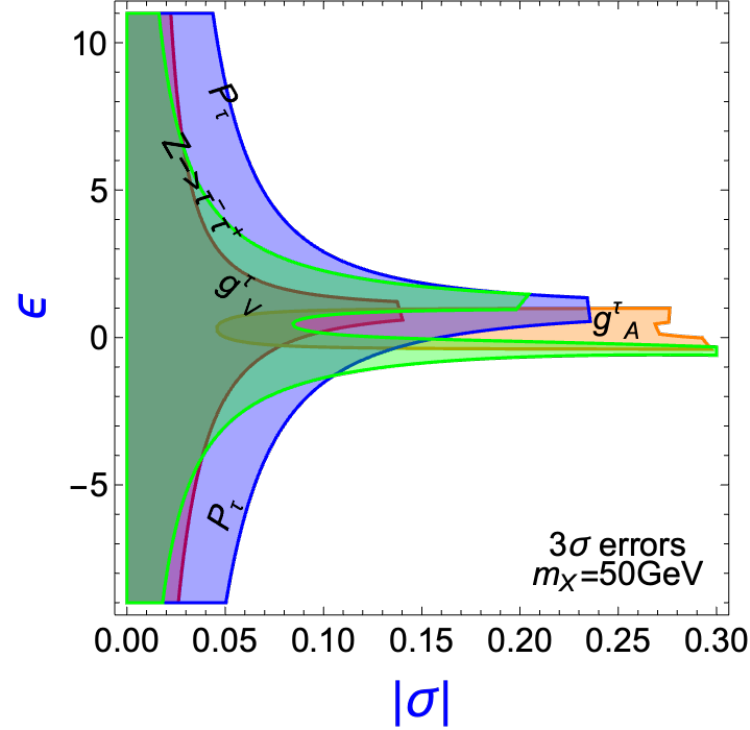}}
	\subfigure[\label{3mx250}]
	{\includegraphics[width=.486\textwidth]{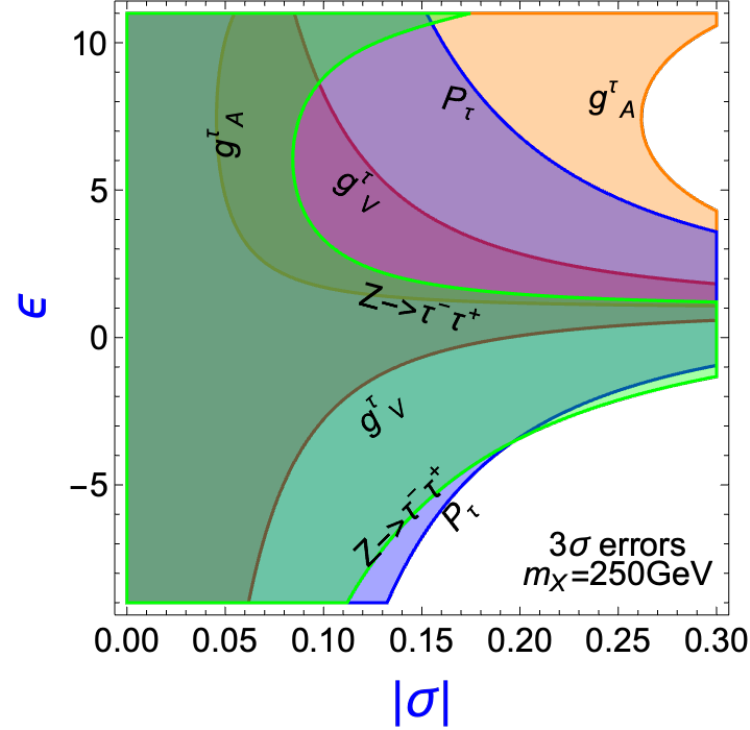}}
	\caption{ The allowed parameter regions in $|\sigma|-\epsilon$ plane with dark photon mass $m_X<m_Z$  and $m_X>m_Z$.  The allowed region by $g_V^\tau$ is in red, $g_A^\tau$ in orange, tau lepton polarization in blue, branching ratio $Br(Z\to \tau^-\tau^+)$ in green,   respectively.  
		(a)  The upper left  panel is for $m_X=50$ GeV within $2\sigma$. 
		(b)  The upper right  panel is for $m_X=250$ GeV within $2\sigma$. 
		(c)  The bottom left  panel is for $m_X=50$ GeV within $3\sigma$. 
		(d)  The bottom right  panel is for $m_X=250$ GeV within $3\sigma$. 
	}
	\label{mass}
\end{figure*}

Because the above four experimental values could both constrain the dark photon model parameters, we  attempt to find the common parameter region satisfying  these constraints simultaneously.   
From Figs.~(\ref{coupling}, \ref{decay}), we found that the mixing ratio parameter $\epsilon$ plays important role in the viable region.  To clearly illustrate the influence of  $\epsilon$ effect, we plot the allowed region in the plane $|\sigma|-\epsilon$ with different dark photon mass  in Fig.~\ref{mass}.
The allowed region by $g_V^\tau$ is in red, $g_A^\tau$ in orange, tau lepton polarization in blue, branching ratio $Br(Z\to \tau^-\tau^+)$ in green,   respectively.

 The upper left  panel is for $m_X=50$ GeV within $2\sigma$ as shown in Fig.~\ref{mx50}.  We found that the regions allowed by vector coupling are totally covered by ones by the tau polarization. The regions by the vector coupling  $g_V^\tau$ exist in $|\sigma|<0.12$.  However,   the ones by the axial-vector coupling  $g_A^\tau$ exist in the opposite direction $|\sigma|>0.12$. And the regions by $g_A^\tau$  can not reconcile with the branching ratio $Br(Z\to \tau^-\tau^+)$. For  small $|\sigma|$
case, there are common spaces satisfying $g_V^\tau$, $Br(Z\to \tau^-\tau^+)$  and tau polarization $P_\tau$ constraints whatever $\epsilon<1$ or  $\epsilon>1$.  Unfortunately, $g_A^\tau$ excluded 
these common spaces. 

 The upper right  panel is for $m_X=250$ GeV within $2\sigma$ as shown in Fig.~\ref{mx250}.  The regions by the axial-vector coupling  $g_A^\tau$ exist in $\epsilon>1$ and  the ones by the $Br(Z\to \tau^-\tau^+)$ exist in the opposite direction $\epsilon<1$.  Similarly, the regions allowed by vector coupling are totally covered by ones by the tau polarization.  There exist common regions meeting  $g_V^\tau$, $P_\tau$ and $Br(Z\to \tau^-\tau^+)$ constraints in small $|\sigma|$ with large negative $\epsilon$.  And  large  $|\sigma|$ with  $\epsilon \in (2, 4)$ could satisfy  $g_V^\tau$, $g_A^\tau$ and $P_\tau$ constraints.  Unfortunately, there are also not common regions satisfying these four bounds simultaneously.  Therefore, we found that dark photon model can not solve these problems within $2\sigma$ errors so that we need to expand the errors into $3\sigma$ case as shown in the bottom of Fig.~\ref{mass}. 
 
 The bottom left  panel is for $m_X=50$ GeV within $3\sigma$ as shown in Fig.~\ref{3mx50}.  We found that the allowed regions  both are expanded, especially for  $g_A^\tau$ and  $Br(Z\to \tau^-\tau^+)$.   Therefore, there exist common regions with $|\sigma| \in (0.04, 0.10)$ and $\epsilon \in (0, 1)$ satisfying these constraints simultaneously. 
 
 The bottom right  panel is for $m_X=250$ GeV within $3\sigma$ as shown in Fig.~\ref{3mx250}. We found that the allowed regions by  $g_A^\tau$ and  $Br(Z\to \tau^-\tau^+)$ are expanded into  left and upper, respectively.   Similarly, it results in  common regions   $\epsilon \in (1, 10)$  satisfying these constraints simultaneously.   
 
 Therefore,  we found that the dark photon with the kinetic and mass mixing within $2\sigma$ could not explain these four constraints at the same time,   vector coupling $g_V^\tau$,  axial-vector coupling $g_A^\tau$, decay branching ratio $Br(Z\to \tau^- \tau^+)$, and tau polarization $P_\tau$.  Fortunately, these bounds within $3\sigma$ can be satisfied simultaneously.

\section{Global fits of $Z\to f \bar f$ processes}
In the above section, we analyze the mixing physical effects for $Z-\tau$ interactions in detail to find the viable parameter regions. Actually, the mixing effects could be extended to the general Z boson interactions with all fermions $f$. 
Consequently, we need to conduct the global fits while considering  all  relevant constraints for $Z\to ff$ as shown in Table.~\ref{table1}. In our analysis, a global fit with Z decay experimental data are given with the non- linear least-$\chi^2$ method~\cite{Zhao:2021xwl}.

\begin{table}[htbp!]
\caption{Experimental data~\cite{pdg} and the best points in the global fit. }\label{table1}
\begin{tabular}{|c|c|c|c|c|c|c|c|c|c|}\hline\hline
Obdervable &Experimental data&Our work $(m_X>m_Z)$\\\hline
$g_V^\tau$ & $-0.0366\pm0.0010$&\multirow{3}{*}{$-0.03814\pm0.00022$} \cr\cline{1-2}
$g_V^\mu$ & $-0.0367\pm0.0023$&\cr\cline{1-2}
$g_V^e$ & $-0.03817\pm0.00047$&\\\hline
$g_A^\tau$ &$-0.50204\pm0.00064$&\multirow{3}{*}{$-0.499784\pm0.000084$}\cr\cline{1-2}
$g_A^\mu $ & $-0.50120\pm0.00054$&\cr\cline{1-2}
$g_A^e $ & $-0.50111\pm0.00035$&\\\hline
$g^u_V $ & $0.266\pm0.034$&$0.19106\pm0.00026$\\\hline
$g_A^u$ & $0.519^{+0.028}_{-0.033}$&$0.499784\pm0.000084$\\\hline
$g_V^d $ & $-0.38^{+0.04}_{-0.05}$&$-0.34629\pm0.00017$\\\hline
$g_A^d$ & $-0.527^{+0.040}_{-0.028}$&$-0.499784\pm0.000084$\\\hline
$Br_{\tau^+\tau^-}$ & $(3.3658\pm0.0023)$\% &$(3.38080\pm0.00091)\%$\\\hline
$Br_{\mu^+\mu^-}$ & ($3.3662\pm0.0066$)\%&$(3.39090\pm0.00091)$\%\\\hline
$Br_{e^+e^-}$ & ($3.3632\pm0.0042$)\%&$(3.39094\pm0.00091)$\%\\\hline
$Br_{c \bar c}$ & $(12.03\pm0.21)$\%&($11.5862\pm0.0074$)\%\\\hline
$Br_{ b \bar b}$ & $(15.12\pm0.05)\%$&($14.8783\pm0.0015$)\%\\\hline
$Br_{(c \bar c+u \bar u)/2}$ & $(11.6\pm 0.6)\%$&($11.5891\pm0.0074$)\%\\\hline
$Br_{(d \bar d+s \bar s+b \bar b)/3}$ & $(15.6\pm0.4)\%$&($14.9390\pm0.0015$)\%\\\hline
$P_\tau$ & $-0.144\pm0.015$&\multirow{3}{*}{$-0.15176\pm0.00090$}\cr\cline{1-2}
$P_\mu$ & $-0.142\pm0.015$&\cr\cline{1-2}
$P_e$ & $-0.1515\pm0.0019$&\\\hline
$P_s$ & $-0.90\pm0.09$&\multirow{2}{*}{$-0.93627\pm0.00022$}\cr\cline{1-2}
$P_b$ & $-0.923\pm0.020$&\\\hline
$P_c$ & $-0.670\pm0.027$&$-0.66708\pm0.00059$\\\hline
$m_W$(GeV) & $80.4028\pm0.0067$&$80.408\pm0.019$\\\hline
$\rho$ & $1.00038\pm0.00020$&$1.00003\pm0.00015$\\\hline
\end{tabular}
\end{table}

Additionally, the mixing effects also affect the W boson mass in terms of oblique parameters STU. Following the procedures~\cite{Cheng:2022aau}, we can obtain 
\begin{eqnarray}
&&\alpha S=4\sigma^2 s_W^2 c_W^2 \frac{1-\epsilon}{1-\tilde r^2}\left(1-s_W^2\frac{1-\epsilon}{1-\tilde r^2}\right)\;,\nonumber\\
&& \alpha T=\frac{\sigma^2 s_W^2}{(1-\tilde r^2)^2}(1-\epsilon)[2\epsilon-\tilde r^2(1+\epsilon)]\;,\nonumber\\
&&\alpha U=\frac{4\sigma^2 s_W^4 c_W^2}{(1-\tilde r^2)^2}(1-\epsilon)^2\;.
\end{eqnarray}
This correspondingly results in the modification for electroweak parameter $\rho=1+\alpha T=0.00038\pm0.00020$. Besides,
the above leads to the correction for W mass as
\begin{eqnarray}
\Delta m_W^2&&=m_Z^2 c_W^2 \left(-\frac{\alpha S }{2(c_W^2-s_W^2)}+\frac{c_W^2 \alpha T}{(c_W^2-s_W^2)}+ \frac{\alpha U} {4s_W^2}\right)\nonumber\\
&&=- m_Z^2 c_W^2 \frac{\sigma^2 s_W^2}{(1-\tilde r^2)(1-t_W^2)}(1-\epsilon)^2\;. \label{mDwmass}
\end{eqnarray}
For the experimental values for W mass, we combine the results from  $80,433.5 \pm 9.4 $ MeV in CDF~\cite{Hays:2022qlw},  $80,360 \pm 16 $ MeV in ATLAS~\cite{ATLAS}, and  $80,377 \pm 12 $ MeV in PDG~\cite{pdg}, to obtain $m_W=80,402.8\pm 6.7$ MeV, which is above than  SM prediction $80,357 \pm 6$ MeV~\cite{pdg}.

As shown in Sec. III, the different experimental bounds can be satisfied within $3\sigma$ error margins, indicating fits within $3\sigma$ errors. In the dark photon model with both kinetic mixing and mass mixing, we choose the parameter ranges as $\epsilon\in [-9,11]$, $\sigma\in [0.0,3]$, and $m_X\in [0,500]{\rm GeV}$. Note that around $m_X \sim m_Z$, there is a resonance implying inadequate fits. This necessitates analyzing the ranges of dark photon mass for $m_X<m_Z$ and $m_X>m_Z$ respectively.

The fit parameters and corresponding correlation matrix are shown in Table.~\ref{table2}. We obtain $\epsilon=-1.37\pm0.46$, $\sigma=0.074\pm0.021$ and 
$m_X=275\pm39$ GeV with $\chi^2/d.o.f=1.23$.
We find that the dark photon mass tends to be in the $m_X>m_Z$ region, aligning with the requirements in Eq.~(\ref{mDwmass}).
\begin{table}[htbp!]
\caption{The fit parameter values for $\epsilon$, $\sigma$ and $m_X$ and the correlation matrix. }\label{table2}
\begin{tabular}{|c|c|c|c|c|c|c|c|c|c|}\hline\hline
parameters &$\epsilon$ &$\sigma$&$m_X$\\\hline
results&$-1.37\pm0.46$&$0.074\pm0.021$&$275\pm39$\\\hline
$\epsilon$&1&0.583&-0.045\cr\cline{1-1}
$\sigma$&&1&0.488\cr\cline{1-1}
$m_X$&&&1\\\hline
\multicolumn{4}{|c|}{$\chi^2/d.o.f=19.75/16=1.23$}\cr\cline{1-4}
\end{tabular}
\end{table}

\section{Conclusion}

We investigate the phenomenology of the dark photon model with both kinetic mixing and mass mixing.
Mass mixing is achieved by introducing an additional Higgs doublet with a vacuum expectation value (vev), which is responsible for breaking both $U(1)_X$ and electroweak symmetry simultaneously.
The interaction between the Z boson and the SM fermions is studied by defining the mixing ratio parameter $\epsilon$, which quantifies the magnitude of both the mass and kinetic mixing of the dark photon.
The two mixing effects will modify the Z boson coupling with the fermions in terms of the dark photon mass $m_X$, kinetic mixing $\sigma$, and mixing ratio $\epsilon$. First, we take the case of tau lepton to conduct the detailed  phenomenological analysis. 
Utilizing experimental results from the vector and axial-vector couplings $g_{V,A}^\tau$, the decay branching ratio $Br(Z\to \tau^- \tau^+)$, and tau lepton polarization in $Z\to \tau^-\tau^+$, we determine the corresponding parameter spaces and analyze the influence of $\epsilon$ by varying its value.
We found that the mixing ratio $\epsilon$ plays an important role in analyzing the Z boson phenomenology.
Furthermore, we attempt to identify common regions that satisfy these four aforementioned constraints for both $m_X>m_Z$ and $m_X<m_Z$.
However, the regions allowed by $g_A^\tau$ and $Br(Z\to \tau^-\tau^-)$ tend to be in opposite directions, rendering it impossible to find viable parameter spaces within a $2\sigma$ error margin. The problem can be resolved within a $3\sigma$ error margin.

The analysis can also be applied to scenarios involving leptons from other generations and even quarks. Therefore, we further extend our study to encompass all fermions by conducting a global fit to determine the most reasonable parameter values, including the kinetic mixing parameter $\sigma=0.074\pm0.021$, mixing ratio $\epsilon=-1.37\pm0.46$, and dark photon mass $m_X=275\pm39$ GeV. Our global analysis indicates a preference for a dark photon mass larger than $m_Z$, which may be beneficial for future investigations into the dark photon model.

\section*{Acknowledgments}
This work was supported by IBS under the project code, IBS-R018-D1.

\end{document}